\documentclass[]{aa}
\usepackage{graphicx}
\usepackage{natbib}
\usepackage[fleqn]{amsmath}
\usepackage[modulo, switch]{lineno}
\usepackage{txfonts}
\def\mh{\,$\mu$Hz}

\def\num{$\nu_\mathrm{max}$}
\def\dnu{$\Delta\nu$}
\def\dnuc{$\Delta\nu_{\mathrm{c}}$}
\def\dnug{$\Delta\nu_{\mathrm{a}}$}
\def\eps{$\epsilon$}
\def\epsc{$\epsilon_{\mathrm{c}}$}
\def\epsg{$\epsilon_{\mathrm{a}}$}
\def\dn1{$\delta\nu_{01}$}
\def\dn2{$\delta\nu_{02}$}

\def\sun{\hbox{$_\odot$}}

\begin{document}
\linenumbers
   \title{Evolutionary influences on the structure of red-giant acoustic oscillation spectra from 600d of \textit{Kepler} observations}

   \author{T. Kallinger\inst{1, 2}
 	  \and
	  S. Hekker\inst{3,4}
	  \and
	  B. Mosser\inst{5}
	  \and
	  J. De Ridder\inst{1}
	  \and
	  T. R. Bedding\inst{6}
	  \and
	  Y. P. Elsworth\inst{4}
	  \and
          M. Gruberbauer\inst{7}
          \and
          D. B. Guenther\inst{7}
	  \and
	  D. Stello\inst{6}
          \and
	  S. Basu\inst{8}
	  \and
          R. A. Garc\'ia\inst{9}
          \and
	  W. J. Chaplin\inst{6}
	  \and
	  F. Mullally\inst{10}
	  \and
	  M. Still\inst{11}
	  \and
	  S. E. Thompson\inst{10}
            }

   \offprints{thomas.kallinger@ster.kuleuven.be}

   \institute{
Instituut voor Sterrenkunde, K.U. Leuven, Celestijnenlaan 200D, 3001 Leuven, Belgium
		\and
Institute for Astronomy (IfA), University of Vienna, T\"urkenschanzstrasse 17, 1180 Vienna, Austria
              	\and
Astronomical Institute ``Anton Pannekoek", University of Amsterdam, PO Box 94249, 1090 GE Amsterdam, The Netherlands
		\and
School of Physics and Astronomy, University of Birmingham, Edgbaston, Birmingham B15 2TT, UK
		\and
LESIA, CNRS, Universit\'e Pierre et Marie Curie, Universit\'e Denis Diderot, Observatoire de Paris, 92195 Meudon cedex, France
		\and
Sydney Institute for Astronomy (SIfA), School of Physics, University of Sydney, NSW 2006, Australia
		\and		
Department of Astronomy and Physics, Saint Marys University, Halifax, NS B3H 3C3, Canada
	\and
Department of Astronomy, Yale University, P.O. Box 208101, New Haven, CT 06520-8101, USA
		\and		
Laboratoire AIM, CEA/DSM-CNRS, Universit\'e Paris 7 Diderot, IRFU/SAp, Centre de Saclay, 91191, Gif-sur-Yvette, France
		\and
SETI Institute/NASA Ames Research Center, Moffett Field, CA 94035, USA
		\and
Bay Area Environmental Research Inst./NASA Ames Research Center, Moffett Field, CA 94035, USA	
}

   \date{Received  / Accepted }

\abstract
{The \textit{Kepler} space mission is reaching continuous observing times long enough to also start studying the fine structure of the observed pressure-mode spectra.}
{In this paper, we aim to study the signature of stellar evolution on the radial and pressure-dominated $l$ = 2 modes in an ensemble of red giants that show solar-type oscillations. }
{We use established methods to automatically identify the mode degree of $l$ = 0 and 2 modes and measure the large (\dnuc ) and small (\dn2) frequency separation around the central radial mode. We then determine the phase shift \epsc\ of the central radial mode, i.e. the linear offset in the asymptotic fit to the acoustic modes. Furthermore we measure the individual frequencies of radial modes and investigate their average curvature.}
{We find that \epsc\ is significantly different for red giants at a given \dnuc\ but which burn only H in a shell (RGB) than those that have already ignited core He burning. Even though not directly probing the stellar core the pair of local seismic observables (\dnuc , \epsc ) can be used as an evolutionary stage discriminator that turned out to be as reliable as the period spacing of the mixed dipole modes. We find a tight correlation between \epsc\ and \dnuc\ for RGB stars and unlike less evolved stars we find no indication that \epsc\ depends on other properties of the star. It appears that the difference in \epsc\ between the two populations becomes smaller and eventually indistinguishable if we use an average of several radial orders, instead of a local, i.e. only around the central radial mode, large separation to determine the phase shift. This indicates that the information on the evolutionary stage is encoded locally, more precisely in the shape of the radial mode sequence. This shape turns out to be approximately symmetric around the central radial mode for RGB stars but asymmetric for core He burning stars. We computed radial mode frequencies for a sequence of red-giant models and find them to qualitatively confirm our findings. We also find that, at least in our models, the local \dnu\ is an at least as good and mostly better proxy for both the asymptotic spacing and the large separation scaled from the model density than the average \dnu . Finally, we investigate the signature of the evolutionary stage on \dn2\ and quantify the mass dependency of this seismic parameter.}
{}

   \keywords{stars: late-type - stars: oscillations - stars: fundamental parameters - instrumentation: Kepler}
\authorrunning{Kallinger et al.}
\titlerunning{Evolutionary influences on the structure of red-giant p-mode spectra}
   \maketitle

\section{Introduction}	\label{sec:intro}

Once a main-sequence star has depleted its supply of H in the core it begins to burn H in a shell surrounding the nearly isothermal He core. With the mean molecular weight increasing, the core is forced to contract under gravity and heats up. To maintain the increasing flow of power from the central regions, the outer layers expand and cool down until the surface opacity dramatically drops due to the shortage of available free electrons from ionised metals that contribute to the H$^-$ opacity. The star has reached the red giant branch (RGB) phase. The core continues to contract, its temperature continues to rise, the luminosity of the burning H shell continues to rise, and the outer layers continue to expand. Because of the drop in surface opacity and the high sub-surface opacity, the subsurface layers become convective in order to maintain the flow of energy out of the star. The star's luminosity and radius increase while its surface temperature remains relatively fixed, i.e., the star climbs the RGB. Early in this phase the convective envelope extends deep into the interior but as the star evolves, the convective region retreats outwards.

Eventually the core is hot enough to ignite He (triple-alpha) burning and the envelope contracts. In lower mass stars, the central densities are high enough that the electron gas is degenerate. The weak temperature dependence of the supporting electron gas pressure leads to a thermal runaway, with the He burning rate increasing rapidly, unconstrained by pressure expansion until degeneracy is removed. The star quickly settles onto the horizontal branch (also called the red clump, RC), burning He in the core and H in a shell, with the latter providing most of the energy. Eventually, the star will exhaust He in the core and the structural changes that occurred earlier will repeat, this time entering the asymptotic giant branch (AGB) phase, with the star moving back up the giant branch. The star is then burning He in a shell surrounding an inert carbon core, and burning H in a shell slightly farther out, being still the dominant energy source.

Even though the general concept of red-giant evolution is well understood \citep[e.g.,][]{ibe68,swe78}, there are a number of outstanding problems, such as mass-loss on the giant branch, deficits in modelling the convective energy transport, and the internal mixing from rotation, that limit our understanding of the chemical evolution of galaxies, with consequences even for cosmology \citep[e.g.,][]{jcd09}.

A problem in addressing these questions is our inability to distinguish the different evolutionary stages of red giants. Having quite different internal structures, they still cover the same region in the Hertzsprung-Russell (HR) diagram and we cannot distinguish them based on their surface properties, i.e., with classical methods like spectroscopy. Asteroseismology offers a solution, being an elegant tool to probe the internal structure of a star by making use of its natural eigenfrequencies, which depend on the physical properties of its interior \citep[e.g.,][]{aerts10}.

All cool stars with a convective surface layer are believed to show high-overtone oscillations of low spherical degree $l$. These so-called solar-type oscillations are intrinsically damped and stochastically excited by the turbulent flux of the near-surface convection. For main-sequence stars, they are pressure (p) modes that are mainly confined to the outer envelope and which follow a  comb-like pattern in the frequency spectrum. The existence of solar-type oscillations in the more evolved red giants is now well established observationally, most recently from CoRoT \citep{bag06} and \textit{Kepler} \citep{bor10}.

However, the dense He cores of red giants complicate the situation, allowing the existence of gravity (g) modes at frequencies that overlap the p-mode frequency spectrum. Red-giant models \citep[e.g.,][]{gue00, dzi01} predict a very dense spectrum of non-radial mixed modes with a g-mode character in the core and a p-mode character in the envelope, whose regular period spacing is directly dependent on the buoyancy frequency in the core region \citep[e.g.,][]{dzi77}. However, g-dominated mixed modes have very high inertias, leading to very small, often unobservable, amplitudes at the surface. From a theoretical perspective, however, some of the mixed modes have substantial amplitudes in the envelope due to resonant coupling between the g- and p-mode cavities \citep[e.g.,][]{jcd04,dup09}. Their mode energy is then concentrated in the envelope giving rise to the surface amplitudes and making them clearly detectable \citep[e.g.,][]{kal08a,rid09,bed10}. For modes with $l>1$, mostly single modes are observable per radial p-mode order with their frequencies following the radial mode pattern. Only for dipole modes ($l$ = 1) the trapping is less selective and more than one mode reaches sufficient surface amplitudes. They appear in groups in between consecutive radial modes forming a characteristic pattern \citep{beck11} with an average period spacing that is smaller than those of the g-dominated mixed modes that are confined to the core. Thanks to the quality of the data obtained with NASA's \textit{Kepler} space telescope for hundreds of red giants, \cite{bed11} (and simultaneously \citealt{mos11b}) found that the observed average period spacings are still a valuable probe for the core properties. They permit a distinction between evolutionary stages of otherwise indistinguishable red giants. 

Here, we concentrate on radial modes and how they are affected by the evolutionary stage of the star. Their frequency pattern is largely defined by a characteristic spacing and a phase shift. From 5-month long CoRoT data, \citet{mos11a} postulated an universal pattern for red-giant oscillations. They measured an average large frequency separation (\dnu ) and indicated that the absolute p-mode frequencies of all red giants are basically a function of \dnu\ and that there is no difference in the \textit{p-mode} spectrum of a RC star and an RGB star if they have the same \dnu . On the other hand, \citet{bed11} found evidence that the frequency spectrum of a red giant with a given \dnu\ is shifted in frequency according to its core properties, i.e. the phase shift of an RGB star differs from that of a RC star, clearly contradicting the universal pattern. We find that this contradiction is caused by different definitions of \dnu . Whereas \citet{mos11a} determined an average \dnu\ of all observed p modes and took into account a gradient of \dnu\ with frequency, \citet{bed11} adopted the large frequency separation that best aligns the strongest radial modes in an \'echelle diagram, which is often different from an average value of all modes. Both methods are potentially sensitive to observational characteristics, such as the signal-to-noise ratio, making it difficult to use them for a high-precision comparison of different stars.
We re-examined the \textit{Kepler} data using a more distinctive ``local'' definition of \dnu , namely the average separation of the three central radial modes. 
Indeed, if we compare stars with the same central \dnu , we find that the central radial modes of core He-burning stars have a smaller phase shift than do stars still burning only H in a shell. Therefore, the three central radial modes can be used to distinguish between the different evolutionary stages of red giants. 

We also investigated the fine structure of the p-mode spectrum and found that the radial mode sequence of RGB stars is approximately symmetric around the central radial mode in contrast to those of RC stars, which is asymmetric or even antisymmetric. Also the small frequency separation between adjacent $l$ = 0 and 2 modes is affected by a star's evolutionary status, leading to a small frequency separation that is on average larger for RC stars than for RGB stars \citep[see also][]{cor12}. We checked the seismic parameters for a scaling with stellar mass and found it to clearly affect the small spacing. Finally, we computed radial mode frequencies for a sequence of red-giant models and found them to qualitatively confirm our findings. 

\section{Observations}	\label{sec:obs}

The \textit{Kepler} space telescope \citep{bor10,koch10} was launched in March 2009 to search for transiting Earth-sized planets in and near the habitable zones of Sun-like stars. \textit{Kepler} houses a 95-cm aperture modified Schmidt telescope that points at a single field in the constellations of Cygnus and Lyra, feeding a photometer with a wide field of view to continuously monitor the brightnesses of about 150\,000 stars. This makes it also an ideal instrument for asteroseismology and the Kepler Asteroseismic Science Consortium\footnote{http://kepler.asteroseismology.com} has been set up to carry this out (see \citealt{gil10} for an overview and first results).

\textit{Kepler} observations are subdivided into quarters, starting with the initial commissioning run (10\,d, Q0), followed by a short first quarter (34\,d, Q1) and subsequent full quarters of 90\,d length. Our studies are based on the long-cadence \citep[29.42 min sampling;][]{jen10} data spanning from Q0 to Q7 with a total of about 27\,200 measurements. Apart from occasional losses of the satellite's fine pointing and scheduled re-orientations of the spacecraft, the $\sim$599 days-long observations were continuous, with in an overall duty cycle of about 93\,\%. The \textit{Kepler} raw data were reduced in the manner described by \citet{gar11} and subsequently smoothed with a triangular filter to suppress residual instrumental long-term trends with time scales longer than about 10\,d. The power density spectra were normalised to recover the full sine-amplitude of an injected signal. Our sample includes a total of 923 red giants showing a clear power excess due to pulsation and has already been used for other studies, such as the investigation of relations between the different seismic \citep{hub10,hub11,mos11c} and granulation \citep{mat11} observables, the seismic determination of fundamental parameters \citep[][hereafter referred to as Paper I]{kal10b}, a comparison of global oscillation parameters derived from different methods \citep{hek10b}, and the analysis of non-radial mixed modes \citep{bed11,mos11d} and how they are used to constrain the core-rotation rate \citep{beck11b}.

\section{The acoustic p-mode spectrum}

The observed oscillation modes of solar-type pulsators are typically high-order acoustic modes. If interaction with g modes can be neglected, the linear, adiabatic, high-order acoustic modes in a spherically symmetric star follow an asymptotic relation for the frequencies with a radial order $n$ and a spherical degree $l$ \citep{van,tas80}:
\begin{equation}
\nu_{n,l} \simeq \Delta \left (n+\frac{l}{2}+\epsilon \right )-d_{0l},
\label{eq:tas}
\end{equation}
where the characteristic (or asymptotic) spacing\footnote{Note that the asymptotic spacing is usually referred to as \dnu\ but in order to keep it clearly distinct from any observed quantity we prefer $\Delta$} $\Delta \simeq \nu_{n+1,l}-\nu_{n,l}$ is the inverse sound travel time across the stellar diameter, essentially scaling with the square root of the mean stellar density \citep[e.g.,][]{bro94}. The phase shift \eps\ is a dimensionless offset and $d_{0l}$ denotes the small frequency separations of non-radial modes relative to radial modes, which is given by $\nu_{n,0} - \nu_{n-1,2}$ in the case of $l=2$ modes. In this first-order approximation, the modes are expected to fall along vertical ridges when folded with $\Delta$ in an  \'echelle diagram \citep{grec83}, with the radial modes shifted by $\epsilon$ times $\Delta$ from the origin and the non-radial modes shifted by $d_{0l}$ relative to the radial modes. This simple concept appears to be quite promising because it suggests that some observed frequency separation can directly be related to physical quantities of a star. However, in reality the situation is more complex. The actual modes fall along curved ridges in the  \'echelle diagram and the departure from the asymptotic relation is a function of frequency (or equivalently of $n$) predominantly determined by the mean stellar structure and atmosphere. Additionally, sharp changes in the acoustic variables, e.g. at the boundary of the He\,\textsc{II} ionisation zone, cause small quasi-periodic variations around the curved mode sequences (see, e.g. \citealt{hou07} and \citealt{mig10} for the Sun and a red giant, respectively, and \citealt{hek11} for red-giant models).

In the classical asymptotic theory, the higher-order correction terms are presented as functions of $\nu$. The radial mode pattern can be written as $\nu(n) = \Delta(\nu) [n + \epsilon]$ with \eps\ fixed, or mathematically equivalent as $\nu(n) = \Delta [n + \epsilon(\nu)]$ with $\Delta$ fixed and preferably $\epsilon(\nu) = \epsilon \, f(\nu)$. We prefer the latter because it allows $\Delta$ to have a single value that is independent of both $l$ and frequency \citep[see also][]{bed11b}. 
This can be developed as follows. The frequencies have to satisfy the Eigenfrequency equation \citep[e.g.,][]{rox00}:
\begin{equation}
 \nu_{n,l} \frac{\pi} { \Delta} = \pi \left( n+\frac{l}{2} \right ) + \alpha_l(\nu) - \delta_l(\nu),
\label{eq:eigenequ}
\end{equation}
where $\alpha_l(\nu)$ is an outer phase shift primarily determined by the mode propagation in the outer layers (including the atmosphere) of the star, and $\delta_l(\nu)$ an inner phase shift determined primarily by the structure of the inner regions. The Eigenfrequency equation can be rewritten so that the phase shift and small separation from Eq.\,\ref{eq:tas} are expressed in terms of phase shift differences (Roxburgh, private communication):
\begin{equation}\label{eq:epsdnu}
\epsilon(\nu) = \frac{1}{\pi} \Bigg (\alpha_0(\nu) - \delta_0(\nu) \Bigg )\\ \\
\end{equation}
\begin{equation}\label{eq:epsdnu1}
d_{0l}(\nu) = \frac{\Delta}{\pi} \Bigg ( \alpha_0(\nu) - \alpha_l(\nu) + \delta_l(\nu) - \delta_0(\nu) \Bigg )\,.
\end{equation}
In fact, the outer phase shifts $\alpha_l(\nu)$ are almost equal for different $l$ and can be taken as independent of $l$. Consequently, $d_{0l}(\nu)$ depends only on the structure of the inner region, more explicitly on the sound-speed gradient near the stellar core. This arises because the small spacing takes the frequency difference of two modes with nearly identical eigenfunction shapes in the outer layers of the star. Therefore, it is sensitive to the structure of the deep interior. The phase shift $\epsilon(\nu)$ has contributions from the outer and inner regions. For the Sun, $\alpha_0(\nu)$ is $\sim$5 times larger than $\delta_0(\nu)$ \citep{rox10} so that the dominant contribution for the solar modes comes from the surface. This is why \eps\ is sometimes referred to as the ``surface term'', which we find not quite appropriate because it clearly has also contributions from the interior. Obviously, the departure from the first order asymptotic relation is given by the variation of $\alpha_0 - \delta_0$ with frequency. 

A problem is to relate the observed large (\dnu ) and small (\dn2) frequency separations and phase shift (\eps ) to the coefficients discussed above, which are not straight-forwardly accessible. The large frequency separation, for example, should ideally be measured from the modes closest to the acoustic cutoff frequency, which presumably follow the characteristic spacing best. But these modes often have very small amplitudes or are unobservable. An average \dnug\ of all observable (radial) modes might be a good alternative \citep[e.g.,][]{whi11b}. Even though it can be determined more accurately and should average out the high-frequency variation along the mode sequence it is potentially sensitive to the mode curvature. For the Sun, for example, the separation of consecutive radial modes varies by almost 6\,\% over the observed range \citep[e.g.,][]{chap99}, with the local separation dropping towards higher frequencies for the lowest radial orders, staying relatively constant around the centre of the power excess hump and slowly increasing towards the highest radial orders. Hence depending on where exactly \dnug\ is measured in this sequence one might end up with values that are significantly different. A careful inspection of many red-giant frequency spectra has shown that the curvature of the radial mode sequence is small \citep[on the order of 1\,\% of \dnu , e.g.,][]{mos11a} but can be quite different in amount or even direction. More interestingly, it appears that the curvature always has a turning point that is near the centre of the power excess, i.e. the curvature is ``in phase'' at this distinctive point. This is also the case for the Sun and other solar-type oscillators. 

To see how well the different choices of \dnu\ represent the asymptotic spacing we investigate theoretical frequencies for a sample of representative models in Sect.\,\ref{sec:model}.

\subsection{Practical implications for red giants}

\begin{figure}[t]
	\begin{center}
	\includegraphics[width=0.5\textwidth]{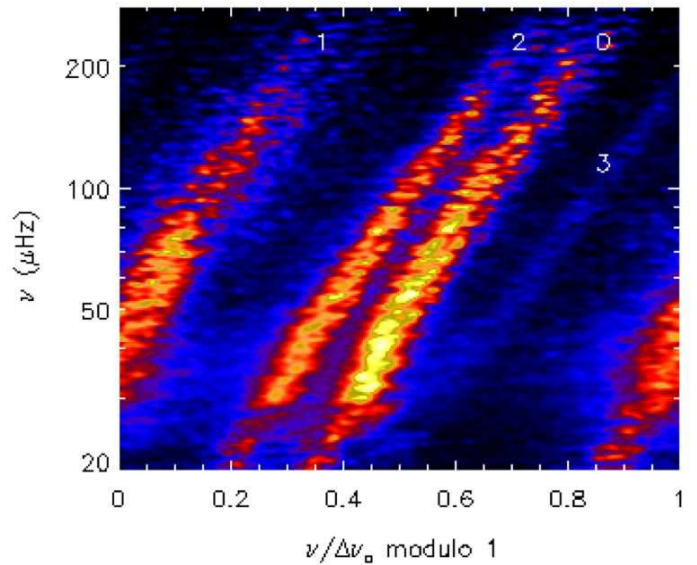}
	\caption{Stacked and normalised  \'echelle diagrams of 923 red giants observed by \textit{Kepler} folded with an average \dnu\ computed according to \citet{mos11a} with the total power colour-coded. The individual spectra are corrected for the granulation signal. Numbers indicate the mode degree.} 
	\label{fig:comp_echelle_global} 
	\end{center} 
\end{figure}

\citet{mos11a} found empirical relations between \eps , $\delta\nu$, and \dnu\ for the red giants observed by CoRoT. Based on these scalings they constructed theoretical oscillation patterns and correlated them to the observed one to  precisely refine their estimate of an average \dnu\ comprising all (except dipole) observed modes. In their approach they assumed that the absolute position of the radial-mode pattern depends only on \dnu, i.e. with \eps\ being a function of only \dnu , regardless of whether the star burns He in the core or still only H in a shell. Consequently, all red giant oscillations should follow a universal pattern and we should not see any difference in the \textit{radial} mode spectrum of RGB and RC stars if they have the same \dnu . This can be seen in Fig.\,\ref{fig:comp_echelle_global}, where we plot normalised  \'echelle diagrams of 923 red giants observed by \textit{Kepler} folded with an average \dnu\ measured using the method of \cite{mos11a} and stacked on top of each other with the resulting total power colour-coded. All spectra show the same pattern with the $l$ = 0 and 2 (and 3) modes falling on top of each other and the phase shift following a simple trend. We note that the scatter along these ridges does not only reflect observational uncertainties or the limited lifetimes of the modes but also the fact that more than one radial order per stars contributes to the total power. As expected the $l$ = 1 ridge is somewhat broadened due to multiple peaks per radial order.

\begin{figure*}[ht]
	\begin{center}
	\includegraphics[width=1.0\textwidth]{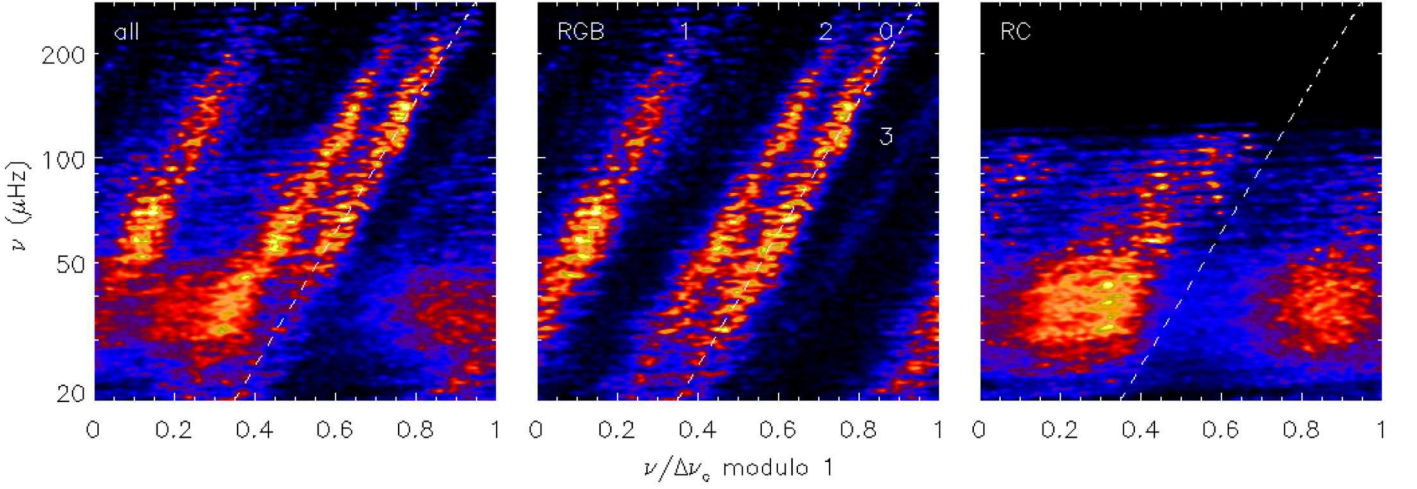}
	\caption{Same as Fig.\,\ref{fig:comp_echelle_global} but now folded with the central \dnu . The left, middle, and right panel show all, only RGB, and only RC stars, respectively, with the identification based on the g-mode period spacings. The white dashed lines represents the empirical relation between \eps\ and \dnu\ from \citet{mos11a}. } 
	\label{fig:comp_echelle} 
	\end{center} 
\end{figure*}

The situation changes if we use a more local \dnu\ instead of an average one to fold the spectra. This can be seen in the left panel of Fig.\,\ref{fig:comp_echelle}. Whereas the observed modes of stars with their oscillation frequencies above about 50\mh\ form a regular pattern with the radial modes (rightmost ridge) following the empirical relation of \citet{mos11a} (white dashed line), stars below about 50\mh\ show a less clear structure \citep[see also][]{hub10}. The latter is the range where we expect both RGB and RC stars \citep[e.g.][]{mig09}. To check if this structural break is due to different evolutionary stages of stars contributing to this diagram, we adopted the identification of \citet{bed11} and indeed found that if we plot only RGB stars (middle panel), the modes form a similar pattern to Fig.\,\ref{fig:comp_echelle_global}, even in the range that is also covered by RC stars (right panel). We note that for RGB stars, modes below about 50\mh\ are barely visible in the left panel because this range is dominated by RC stars, which clearly outnumber the RGB stars. 

The different positions of the mode ridges in Fig.\,\ref{fig:comp_echelle} indicate that RC stars have a phase shift that differs systematically from that of RGB stars. This was already noted by \citet{bed11} but their measurements of \eps\ were not accurate enough to serve as an unambiguous indicator for the stars' evolutionary stage. This might be caused by their specific choice to determine \dnu\ over all observed modes.

\subsection{The central large frequency separation}

To follow up on this interesting difference we consider a more distinctive definition of the large frequency separation, namely the average separation of the three radial modes around the centre of the pulsation power excess (\dnuc ), as introduced by \citet{kal10a} and Paper I.

First we fitted a global model to the observed power density spectra to determine the granulation background signal. The model consists of a superposition of white noise, the sum of super-Lorentzian\footnote{The adopted background model is a modified Lorentzian with the exponent equal to 4 (instead of 2) and is therefore called ``super-Lorentzian''} functions, and a Gaussian to approximate the power excess due to pulsation. The latter component provides a good estimate for the centre of the pulsation power excess, the so-called frequency of maximum oscillation power or \num . For more details about the fitting we refer to Paper I and \citet{mat11}.

\begin{figure}[t]
	\begin{center}
	\includegraphics[width=0.5\textwidth]{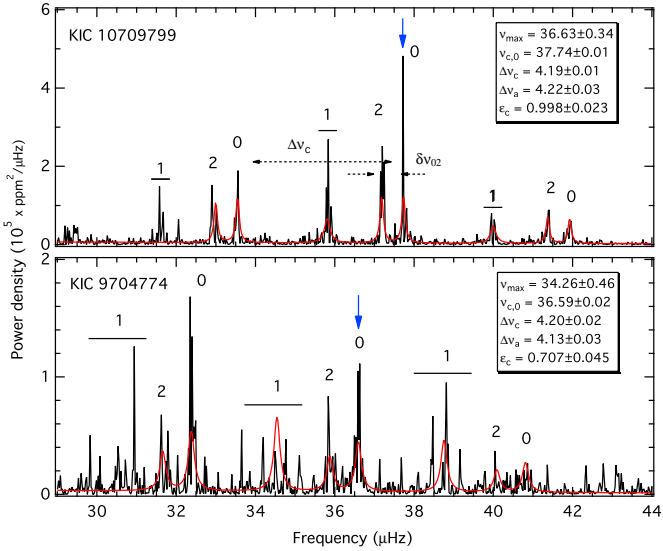}
	\caption{Power density spectra of an RGB (top) and RC (bottom) star with almost identical \dnuc\ but with the central radial modes ($\nu_{c,0}$, indicated by blue arrows) at significantly different frequencies (resulting in different $\epsilon_c$). The red lines indicate the best fit model to determine the p-mode spacings, \dnuc\ and $\delta\nu_{02}$. Numbers give the mode degree.} 
	\label{fig:fourier} 
	\end{center} 
\end{figure}

In the next step, we used the global model fit without the Gaussian component and fitted on top of that a sequence of Lorentzian profiles. The sequence covers three radial orders of $l$ = 0 and 2 modes around \num , where the mode frequencies are parameterised by the frequency of the central radial mode ($\nu_{c,0}$) and a large (\dnuc ) and small (\dn2) spacing. A single linewidth parameter was assumed for the six profiles, which might not be true in reality, but has no impact on the determination of the spacings and significantly stabilises the fit. To account for the power of the intervening p-dominated mixed dipole modes, two additional profiles were used, where the linewidth was fitted independently of the linewidth of the $l$ = 0 and 2 modes. Fig.\,\ref{fig:fourier} illustrates the power density spectra of RGB (top) and RC (bottom) stars with almost identical \dnuc , along with the best fitting models (red lines). In the case of the RGB star with very closely spaced dipole modes (that are almost unresolved), the two additional profiles give a good estimate of where to expect the pure $l$ = 1 p modes, parameterised by $\delta\nu_{01}$, which is the separation from the midpoint of adjacent radial modes \citep[e.g.][]{bed10}. For the RC star, whose individual dipole modes are well separated, the two additional profiles give no useful ``physical'' information. Their purpose here is to represent the integrated power of the dipole modes to stabilise the fit for the $l$ = 0 and 2 modes. 

To further reduce the number of free parameters for the fit we parameterised the heights of the individual profiles with a Gaussian envelope for all modes of a given degree. The resulting three Gaussians utilise a single central frequency and width parameter but individual heights for each mode degree. This resulted in a total of 11 free parameters to fit (compared to 24 if everything is left free). Instead of the Markov-Chain Monte Carlo (MCMC) algorithm used in Paper I, we applied a Bayesian inference tool called \textit{MultiNest} \citep{feroz09}. The big advantage of the nested sampling algorithm is that it provides the posterior distributions for the parameter estimation as well as the model evidence, without the need for multiple runs (as it would be necessary for a MCMC algorithm) and thereby significantly outperforms any other fitting method in terms of efficiency and robustness. With a reliable statistical measurement in hand, telling us how well a given model represents the observations, we could investigate different parameterisations of the eight Lorentzian profiles. 
Usually, more complex models (i.e., with more free parameters) tend to fit some data better than less complex models but in a Bayesian analysis a model is assigned a penalty for its complexity and needs to fit the data considerably better to get a higher model evidence than a less complex model. Fitting different complex models (e.g., with independent mode height and/or linewidth parameters) to a number of typical red-giant spectra, we found that the above model represents the observations best. Adding further parameters (i.e., making the model more complex) did not improve the fit (i.e., the model evidence) for the present observations.

During the fit, the centre and width of the mode height envelopes were allowed to vary over the ranges $\nu_\mathrm{max}\pm\sigma_g$ and $\sigma_g \pm0.5\sigma_g$, respectively, where $\sigma_g$ corresponds to the width of the power excess, as measured from the background fit. 
The linewidth parameters were sampled in a range that corresponds to a mode lifetime of 10 to 100 days for the $l$ = 0 and 2 modes, and 2 to 100 days for the $l$ = 1 modes. Note that a lifetime of 100 days corresponds to a linewidth that is about twice the frequency resolution of the observations. The individual envelope heights were sampled between 0 and twice the power density of the highest peak in the frequency range of interest. A critical parameter is the central frequency because the central mode must be a radial mode in order for the fitted spacings to be meaningful. We allowed $\nu_{c,0}$ to vary in the range $\nu_\mathrm{max}\pm0.55$ times an initial guess for \dnu\ taken from previous analysis, as in Paper I. Since we already have good estimates for \dnu\ and \dn2\ from previous work, we can now be more restrictive for their sampled parameter ranges than in Paper I. The spacing parameters were allowed to vary between 0.8 and 1.2 times the initial guess values (mostly taken from Paper I), resulting in a significantly lower number of misidentified stars. The algorithm automatically identified the central radial mode (and therefore provides a mode degree identification for all other modes) and determined the central ($\pm$1 radial order around \num ) large and small frequency separations for about 80\,\% of the stars in our sample. For the remaining stars, we iteratively adjusted the parameter ranges by hand to achieve a correct identification and finally checked each single fit for its plausibility. We determined the most probable parameters and their 1$\sigma$ uncertainties from the marginal distributions of the probability density delivered by \textit{MultiNest}. The large and small frequency separations were determined to within 0.4 and 4.5\,\%, respectively, for 50\,\% of our sample and to better than 1 and 10\,\%, respectively, for 95\,\% of the stars. The frequency of the central radial mode was measured to within 16\,nHz for 50\,\% of the stars and better than twice the frequency resolution ($\sim$39\,nHz) for almost the entire sample.

\subsection{The phase shift of the central radial mode} \label{sec:eps}

For the remaining analysis we consider the asymptotic relation for the three central radial orders alone. For radial modes, the relation is 
\begin{equation}\label{eq:eps_c}
\nu_{c,0} = \Delta\nu_c\,(n+ \epsilon_c'). 
\end{equation}
Knowing $\nu_{c,0}$ and \dnuc\ we can easily determine the phase shift of this mode as 
$\epsilon_c' = (\nu_{c,0} / \Delta\nu_c)$ modulo 1. 
For the Sun, $\epsilon$' is small (i.e. closer to 0 than to 1) and per convention \eps\ is defined as $\epsilon$'+1 to obtain the correct radial order of the solar p modes. 
\cite{whi11b} showed that \eps\ changes smoothly from the Sun to the red giants. Hence, to obtain values of \eps\ for the red giants that are consistent with the solar value we define,

\begin{equation}
\epsilon = \left\{ \begin{array}{rl}
 \epsilon' + 1 &\mbox{ if $\epsilon' < 0.5$ and $\Delta\nu >$ 3\mh} \\
  \epsilon' &\mbox{ otherwise}
       \end{array} \right.
\end{equation}

\begin{figure}[t]
	\begin{center}
	\includegraphics[width=0.5\textwidth]{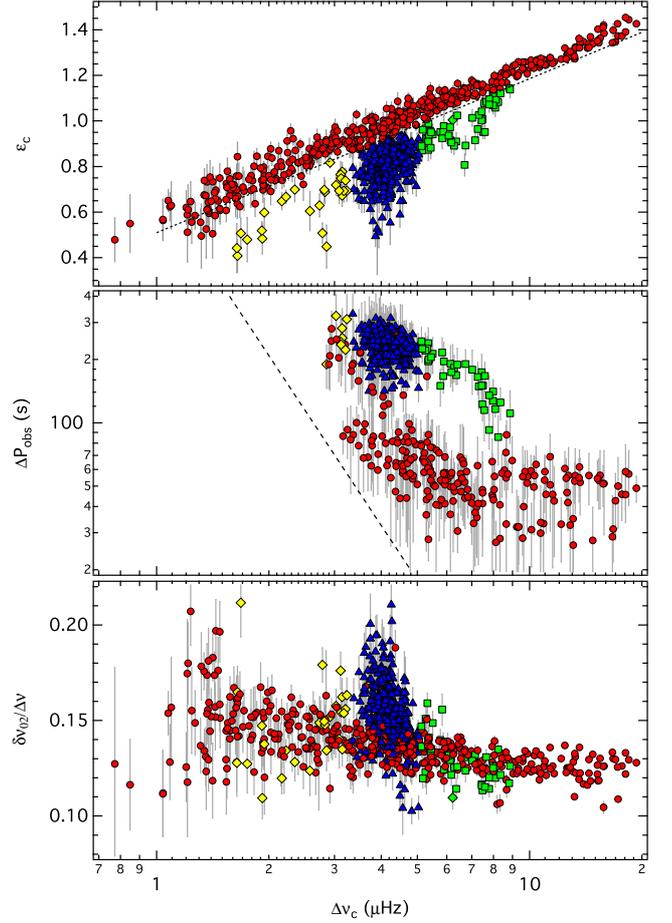}
	\caption{Various seismic parameters as a function of \dnuc . \textit{Top panel:} The phase shift \epsc\ of the central radial mode. The stars divide into two clear groups, with red circles indicating H-shell burning stars. The remainder are core He-burning stars divided into stars that presumably belong to the red clump (blue triangles), the secondary clump (green squares), or the asymptotic giant branch (yellow diamonds). The dotted line is an arbitrary limit to divide the different groups. \textit{Middle panel:} The average period spacing of mixed dipole modes with the classification from the top panel. The dashed line indicates the minimum separation that can be measured between two peaks around $\nu_{c,0}$ in a 600\,d-long time series. \textit{Bottom panel: } Ratio between central small and large frequency separation. Colours are only given in the online version.} 
	\label{fig:eps} 
	\end{center} 
\end{figure}

The top panel of Fig.\,\ref{fig:eps} illustrates the phase shifts of the central radial modes for our sample of 923 red giants as a function of \dnuc\ (note the logarithmic abscissa). Error bars indicate the uncertainties determined from the uncertainties in \dnuc\ and $\nu_{c,0}$. The stars clearly divide into different groups. We subdivided the stars to fall above and below a certain limit (dotted line), which was initially given by a linear fit (in the \epsc\ - $\log \Delta\nu_c$ plane) to a number of stars from the upper population and arbitrarily shifted by -0.05 in \epsc . This works fine for stars with \dnuc\ above about 3\mh . For the stars below 3\mh , the larger uncertainties blend the different populations and we only assigned those to the upper group that are within 2$\sigma$ of the initial (unshifted) fit. The stars above the limit presumably belong to the ascending giant branch and are indicated by red circles. The stars below the limit are presumably core He-burning stars, which we further subdivide into the following groups. The red clump stars (blue triangles) are expected to cover only a narrow range in \dnuc. They are low-mass stars that suffered from electron degeneracy in their RGB phase and ignited helium burning in a flash once the core mass had reached a critical limit. This common core mass means that they all produce about the same amount of energy, which explains why they span a very narrow range of luminosities and radii (and due to a similar mass also \dnuc ). Stars with \dnuc\ above about 5\mh\ (green squares) are secondary-clump stars \citep{gir99} that are too massive to have undergone a helium flash and ignite He in a non-degenerative way, which allows for a range of core masses and therefore of luminosities and radii (and \dnuc ). There is no counterpart to the secondary-clump stars at the low-mass end of the RC below about 3.2\mh\ (yellow diamonds) because stars with an even lower mass than RC stars are not yet old enough to having reached this late stage of red-giant evolution. We identify them as stars that presumably have already left the RC to climb the AGB. However, \epsc\ is quite uncertain in this domain and some of the AGB stars are likely misidentified.

To check if our distinction of the evolutionary stages based on the phase shift of the central radial mode does actually make sense, we followed the idea of \cite{bed11} and \cite{mos11b} of determining the average period spacing of the mixed dipole modes. To do so, we used our mode identification as a template to cut out $\pm$2 radial orders (around $\nu_{c,0}$) of $l$ = 0 and 2 modes from the power density spectra. The remaining signal should then only include dipole modes (and $l$ = 3 modes in some cases). We then split the spectrum (now converted to periods) into four consecutive parts (two above and two below $\nu_{c,0}$), each convering the frequency range between a radial mode and the next $l$ = 2 mode, and computed the highest peak in the autocorrelation function (above twice the frequency resolution) of each part, provided there were at least two well-separated peaks present exceeding nine times the background level. The final average period spacing and its uncertainty were determined from the individual measurements and their standard deviation, respectively. We were able to measure average period spacings for about 2/3 of the stars in our sample. They are given in the middle panel of Fig.\,\ref{fig:eps}, colour-coded according to the classification based on the phase shift. 
\cite{jcd11} argued that He burning causes a convective core where the g modes are excluded, resulting in a period spacing in RC stars that is greater than those of their RGB counterparts. According to \cite{bed11}, the p-dominated dipole modes are separated by about 50 and 200 seconds in RGB and RC stars, respectively. The secondary clump stars cover roughly the range in between, but at higher \dnuc\ than the RC stars. We see that our identification of the evolutionary stage based on \epsc\ is correct for most of the stars in the sample. About half of the stars whose classifications do not coincide are located in the region where it is difficult to separate RGB from RC stars in the \epsc\ - \dnuc\ diagram (around the dotted line in Fig.\,\ref{fig:eps}). For the remainder, it is likely that we measured multiples of the actual period spacing because the individual dipole modes were only poorly resolved. For stars with \dnuc\ below about 3\mh\ the observations are not yet long enough to resolve the individual dipole modes and significantly longer observations are needed to follow stellar evolution towards the tip of the RGB. This clearly shows that the phase shift of the central radial mode is as reliable as the average period spacing of p-dominated dipole modes for determine the evolutionary stage of red giants with \dnuc\ $\gtrsim$ 3\mh\ and can even provide at least a statistical criterion for stars with \dnuc\ $\lesssim$ 3\mh\ to distinguish between RGB and AGB stars, which is not yet possible with the period spacing.

\subsection{The central small frequency separation}

We have also measured the average frequency separation between the three radial and $l$ = 2 modes around $\nu_{c,0}$. We show the ratio $\delta\nu_{02}/\Delta\nu_c$ in the bottom panel of Fig.\,\ref{fig:eps}. The small spacing measures the frequency difference of two modes that have nearly identical eigenfunction shapes in the outer layers of the star and is therefore sensitive to anything that would change the slope of the sound speed in the deep interior. For stars on the main sequences this includes the slow synthesis of a He core and \dn2\ can be used as an age indicator. It, however, fails as an indicator of age when the core is not undergoing significant evolutionary changes, which occurs during the sub-giant phase where the core is still relatively large. On the giant branch, the core is so centrally concentrated that \dn2\ can once again be used as a measure of evolution as it can be sensitive to the state of the H burning shell \citep[e.g.,][]{whi11b}. The cores of RGB and RC stars, being slightly different, will be affected differently. And this is in fact what we see. During RGB evolution, \dn2\ increases continuously to come back to a value after ignition of core He burning in the RC that is on average about 10\,\% larger than those of their RGB counterparts. However, different physical phenomena occurring during different evolutionary stages have similar effects on \dn2\ and no clear distinction between them can be made.

\begin{figure}[t]
	\begin{center}
	\includegraphics[width=0.5\textwidth]{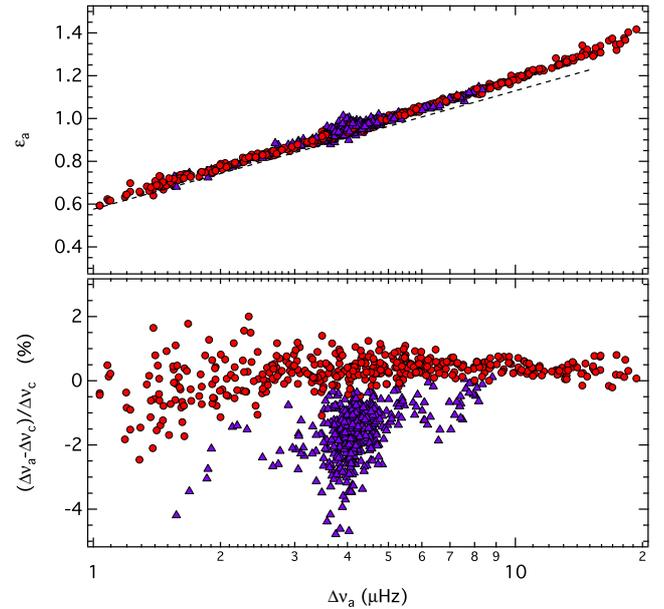}
	\caption{\textit{Top panel:} The phase shift \epsg\ of the central radial mode as derived from the average \dnug\ with the $\epsilon(\Delta\nu)$ function from \citet{mos11a} superimposed (dashed line shifted by -0.05 for better visibility). The classification is taken from Fig.\,\ref{fig:eps} with red circles indicating H-shell burning stars and blue triangles core He-burning stars (including secondary clump and AGB stars). \textit{Bottom panel:}  Relative difference between central and average \dnu . Uncertainties are typically well below 1\,\%.} 
	\label{fig:eps_diff} 
	\end{center} 
\end{figure}

\subsection{Average vs. central large frequency separation}

To investigate the differences between Fig.\,\ref{fig:comp_echelle_global} and \ref{fig:comp_echelle}, we computed the phase shift for the central radial mode with respect to the average \dnug\ from  \citet{mos11a} and plot them in the top panel of Fig.\,\ref{fig:eps_diff}. Obviously, the information on the evolutionary stage is hidden in the \epsg\ - \dnug\ diagram indicating that the evolutionary effect on the radial mode pattern is a local phenomena encoded in the fine structure of the mode sequence. To check if the approach of \citet{mos11a} somehow artificially tunes \dnug\ so that \epsg\ is forced to fall along a perfect relation we re-computed our central \dnu\ for a broader range of modes and find that this is clearly not the case. Now fitting five instead of three radial orders, our central \dnuc\ is still not a global value, but the effect of stellar evolution on \epsc\ fades with the different populations in Fig.\,\ref{fig:eps} blending into each other. Interestingly, the difference between the average and central \dnu\ is only significant for core-He burning stars as shown in the bottom panel of Fig.\,\ref{fig:eps_diff}. Whereas there is no or only a marginal difference for RGB stars, core He-burning stars have a \dnug\ that is on average about 1.5\,\% smaller than their \dnuc . We note that this opens up the possibility to identify the evolutionary stage of a red giant without the need to determine a phase shift by comparing the different large frequency separations.

\subsection{Radial mode curvature of RGB and RC stars}

\begin{figure}[t]
	\begin{center}
	\includegraphics[width=0.5\textwidth]{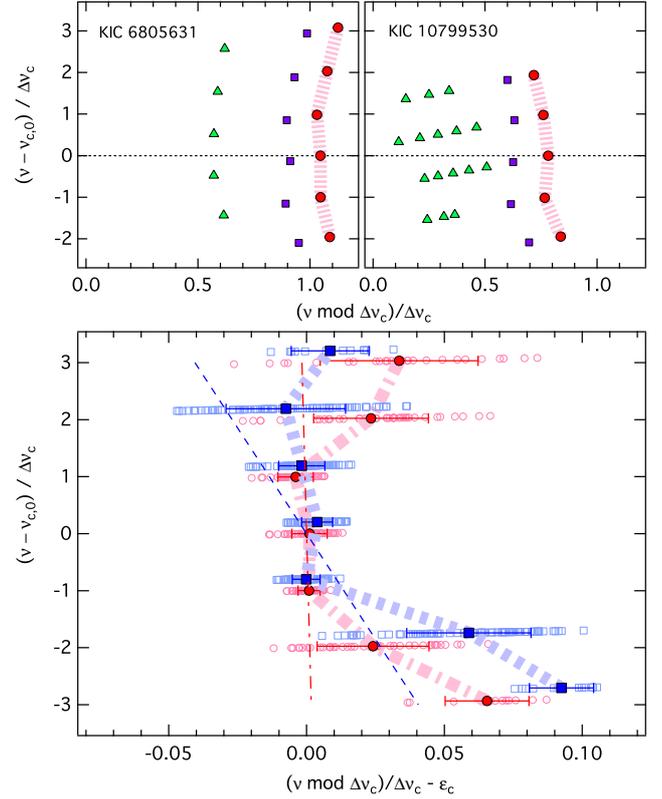}
	\caption{\textit{Top panels:} Normalised  \'echelle diagrams of an RGB (left) and RC (right) star with $l$ = 0, 1, and 2 modes indicated by red circles, green triangles, and blue squares, respectively. \textit{Bottom panel:} Radial mode sequence of 48 RGB (open light-red circles) and 89 RC (open light-blue squares) stars with \dnuc\ ranging from 4 to 5\mh . The abscissa is corrected for the phase shift of the central radial mode. Average values per radial order are given by blue-filled squares (RC stars) and red-filled circles (RGB stars) connected by broad dashed and dash-dotted lines, respectively. Error bars indicate the rms scatter. Thin lines give the slope of a linear fit to the mode sequences, which differ by about 1.5\,\%.} 
	\label{fig:curv} 
	\end{center} 
\end{figure}

We find that the difference between the average and central \dnu\ originates from different shapes of the mode sequences. When plotting individual mode frequencies of an RGB and RC star with similar \dnuc\ in a normalised  \'echelle diagram (top panels of Fig.\,\ref{fig:curv}), we find that the radial modes follow differently shaped sequences. Whereas the radial mode sequence of the RGB star (left panel) is ``C'' shaped (i.e. approximately symmetric around the central radial mode), the mode sequence of the RC stars (right panel) is more ``S'' shaped (i.e., antisymmetric around the central radial mode). In the case of the symmetric mode sequence, \dnu\ determined from the three central radial modes gives a very similar value than the average of all modes. Not so for the antisymmetric sequences, where we find a significant difference between the average and central \dnu .

To test if this is a common phenomenon for the different red-giant populations, we need to extract individual mode frequencies for a larger sample of stars. Reliable peakbagging for a large number of red giants is a complex task that is not yet solved satisfactorily and is also beyond the scope of this paper. However, since we are only interested in an average mode sequence for an ensemble of stars, we adopted a very fast and simple approach. Using the template
\begin{equation}\label{eq:radial_template}
\nu_{n,0} = \nu_{c,0} + \Delta\nu_c\, \left (n'+ \frac{\alpha}{2} n'^2 \right ) 
\end{equation}
with $\alpha \simeq 0.008$ \citep{mos11a} and $n'$ being the radial order relative to the central radial mode, we could automatically estimate the positions of the radial modes. The fit to determine the various frequency separations also provides an estimate for the average linewidth of $l$ = 0 and 2 modes, which we find it to roughly scale as $\Gamma \simeq 0.026$ + $0.12 \log \Delta\nu$ (with $\Gamma$ and \dnu\ in \mh ). This is in good agreement with what was found by \cite{cor12}. We then determined the individual mode frequencies as the centre of area $\pm3\Gamma$ around the initial guess, provided there was a peak present that exceeded nine times the background level.

The result is shown in the bottom panel of Fig.\,\ref{fig:curv}, where we plot the radial mode frequencies of 48 RGB and 89 RC stars, with \dnuc\ ranging from 4 to 5\mh , in a normalised \'echelle diagram. To account for the varying \epsc\ in this range, the abscissa is corrected for the phase shift of the central radial mode (as determined in Sect.\,\ref{sec:eps}). The scatter at the central mode reflects the uncertainties of our peakbagging approach and is, with typically twice the frequency resolution, significantly larger than if we would have fitted proper Lorentzian profiles. Obviously, the mode curvature is not only determined by the evolutionary stage of the star but the general trend is clear. RGB stars have a radial mode sequence that is approximately symmetric around the central mode and RC stars show a strongly asymmetric or even antisymmetric mode sequence. The slope of a linear fit to the average mode sequences gives the difference between an average (comprising all modes) and central \dnuc , with a vertical line indicating no difference. Whereas the difference is negligible for RGB stars, the RC stars show a significant difference, with the average value being on average about 1.5\,\% smaller than \dnuc . This rather small difference translates into a significant change in \epsc\ of typically 0.14 at \dnuc\ = 4.5\mh , which is enough to move a RC star into the RGB population in the \epsg\ -- \dnug\ diagram.

\section{Red-giant models}

\begin{figure}[t]
	\begin{center}
	\includegraphics[width=0.5\textwidth]{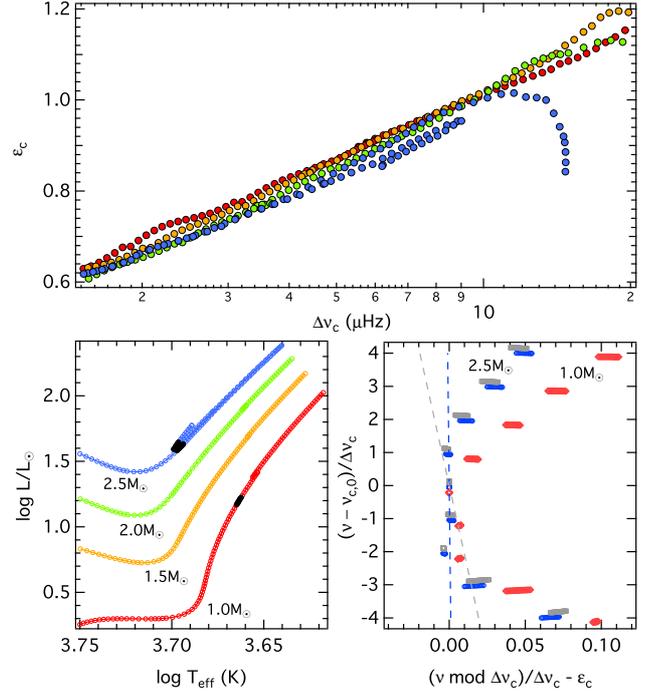}
	\caption{\textit{Top panel:} Phase shift of the central radial mode for a sequence of 1.0, 1.5, 2.0, and 2.5\,$M$\sun\ ascending RGB models. Descending RGB and core He-burning models are shown only for the 2.5\,$M$\sun\ track. \textit{Bottom left panel:} The models from the top panel shown in a theoretical HR-diagram with the models shown in the bottom right panel indicated with black symbols. \textit{Bottom right panel:} Normalised \'echelle diagram showing the radial modes of 1.0 (red diamonds) and 2.5\,$M$\sun\ (blue circles) RGB models and 2.5\,$M$\sun (grey squares) RC models. The blue and grey lines indicate a linear fit to the RGB and RC mode sequences, respectively. The grey and red symbols are vertically shifted for better visibility. Colours are only given in the online version.} 
	\label{fig:model_eps} 
	\end{center} 
\end{figure}

To compare the different choices of \dnu\ to more physical quantities and see whether we can reproduce the observed evolutionary effects on the local fine structure of radial modes, we constructed a sequence of red-giant models.
They were computed with the Yale Stellar Evolution Code \citep[YREC;][]{gue92,dem08} for a near solar composition and calibrated mixing length parameter (Z = 0.02, Y = 0.27, $\alpha_\mathrm{MLT}$ = 1.8) assuming standard solar mixture \citep{grev96}. More details about the constitutive physics may be found in, e.g., \citet{kal10c} and references therein. Like most other stellar evolution codes, YREC is not able to evolve models with a mass below about 2.3\,$M$\sun\ through the helium flash. We therefore had to restrict our analysis to low-mass models on the ascending giant branch and we  only evolved a 2.5\,$M$\sun\ model beyond the ignition of He burning until the exhaustion of He in the core in the secondary clump. The giant branch part of the evolutionary tracks are illustrated in the bottom left panel of Fig.\,\ref{fig:model_eps}. The adiabatic pulsation spectrum of every model was calculated using Guenther's nonradial nonadiabatic stellar pulsation program \citep{gue94}.

\subsection{Average vs. central \dnu\ in red-giant models} \label{sec:model}

Ideally, the large frequency separation should be measured from the highest radial order modes, which follow the asymptotic spacing best. These modes, however, are usually not accessible or are uncertain due to their small amplitudes. An important question, therefore, is how well the various measurements of \dnu\ compare to the asymptotic spacing and to what extent they can be used in the asymptotic relation. Unfortunately, we cannot measure the sound travel time across a real star, and therefore the asymptotic spacing, directly but we can at least ask how accurately the measured \dnu\ follows it for stellar models.

For a reasonable comparison it is important to extract the average and central \dnu\ from the models in a similar way as from the observations. We follow the approach of \citet{whi11b}, where only the modes within a specific range around \num\ were used to measure \dnu , with \num\ determined from a scaling relation \citep[e.g.,][]{kje95}. We could compute the average frequency separation of the modes within the specific range, but this would cause unphysical jumps as modes at the top and bottom fall in and out of this range when \num\ varies from one model to the next. A more appropriate approach is to compute a weighted average of all modes, with the weights given by a Gaussian centred on \num\ \citep[as done by][]{whi11b}. The frequency range, and therefore the number of modes for which \dnu\ is determined, can be controlled by the width of the Gaussian. For the average \dnu\ we chose FWHM = 0.74\num$^{0.82}$, which is a good estimate for the actual width of the observed power excess \citep[e.g.,][]{mos10}. Things are more difficult for the central \dnu , since only a very narrow range has to be considered. We found reasonable results with the FWHM being equal to twice the \dnu\ scaled from the model mean density, essentially giving weight to only the three central modes.
\begin{figure}[t]
	\begin{center}
	\includegraphics[width=0.5\textwidth]{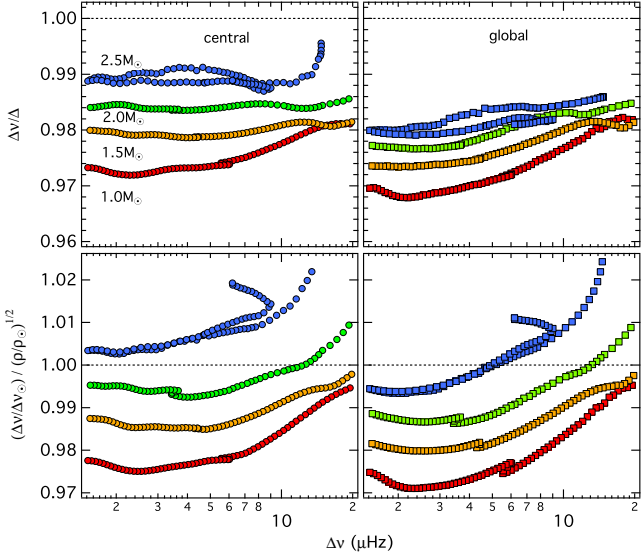}
	\caption{\textit{Top panels}: Ratio of the central (left) and average (right) \dnu\ to the asymptotic frequency separation $\Delta$ for a sequence of red-giant models. Dashed lines indicate unity. \textit{Bottom panels}: Same as top panels but in reference to the square root of the mean model density, where we adopt a \dnu\sun\ of 134.88\mh\ (Paper I).} 
	\label{fig:model_dnu} 
	\end{center} 
\end{figure}

The ratios of the central and average \dnu\ to the corresponding asymptotic spacing (as determined from the models' acoustic radius) are illustrated in the top panels of Fig.\,\ref{fig:model_dnu} for models ranging from 1.0 to 2.5\,$M$\sun . We see that both central and average \dnu\ underestimate the asymptotic spacing by typically 1 to 3\,\%, but interestingly there is a clear trend to approach $\Delta$ with increasing mass. This is because the mode sequence of low-mass stars is more curved than those of high-mass stars, i.e. the spacing of the ``unobserved'' highest radial order modes differs stronger from a central or average \dnu\ (see below). Even though the central \dnu\ is more sensitive to mass, i.e. it covers a larger range of the ratio $\Delta\nu/\Delta$, it is almost always closer to $\Delta$ than is the average \dnu . This is especially true for the red-giant regime. If we go to less evolved sub-giant and main-sequence stars with \dnu\ above 20\mh\ (which are not shown in Fig.\,\ref{fig:model_dnu}), the central and average \dnu\ tracks converge and eventually fall on top of each other. One might expect that the deviation between ``observed'' and asymptotic \dnu\ is at least partly due to the so-called near-surface effect, which is associated with deficiencies in modelling the upper stellar layers \citep[e.g.,][and references therein]{gru11}. But in fact taking this into account would add another $\sim$0.8\,\% deviation to the ratio $\Delta\nu/\Delta$ (at least if we assume the near-surface effect for red giants to be solar-like in amount and direction). 

The situation is slightly different when we compare the central and average \dnu\ with the square root of the mean model density in the bottom panels of Fig.\,\ref{fig:model_dnu}. The ratio $\Delta\nu/\rho^{1/2}$ (in solar units) compares well to $\Delta\nu/\Delta$ for the low-mass models, but due to an even more pronounced mass dependency, it approaches unity faster and even overestimates the reference value for high-mass models. Again, the central \dnu\ appears to be a slightly better proxy for the \dnu\ scaled from the model density than the average \dnu\ (except for high-mass stars, which comprise only a very small fraction of our red-giant sample). We note that our findings are fully consistent with those of \citet{whi11b}, who performed a similar comparison but using ASTEC models with somewhat different input physics.

Even though we cannot investigate the difference between the RGB and RC population for low-mass models, the secondary clump models do not show any significant difference from their RGB counterparts in the comparison with the asymptotic spacing. Only in the comparison with the model density do we find the secondary clump models to differ from the RGB models, indicating that the relation between model density and large frequency separation (i.e., the \dnu\ scaling relation) is different for RGB and core He-burning stars, as already suggested by \cite{mig11}.

\subsection{The phase shift and mode curvature in red-giant models}

Finally, we tested whether our models can reproduce the observed evolution of the phase shift and the radial mode curvature. The top panel of Fig.\,\ref{fig:model_eps} illustrates the phase shift \epsc\ of the models' central radial mode, determined in exactly the same way as from the observations. Apart from a systematic shift towards lower \epsc\ (which is likely due to the near-surface effect), the RGB models follow the same linear trend with \dnuc\ as the observations, with negligible influence from mass. More interesting here is the ``evolution" of \eps\ beyond the ignition of core He burning of the 2.5\,$M$\sun\ models. At the tip of the giant branch the stellar envelopes start to contract and \epsc\ increases again with increasing \dnuc\ but at a slightly lower rate than the RGB models were previously decreasing. When the models settle in the secondary clump at \dnuc\ around 9\mh , \epsc\ is about 0.05 smaller than those of their RGB counterparts. This is about the same difference we observe for many red giants in the corresponding \dnuc\ range. Note that this simple monotonic behaviour of \epsc\ is not found for less evolved sub-giant and main-sequence stars, where stellar mass has a large influence on \eps\ \citep[e.g.,][]{whi11b}. This is already the case for the 2.5\,$M$\sun\ models with \dnuc\ above $\sim$10\mh . However, they evolve very fast and are therefore quite rare so that we do not find a single one in our sample.

The different mode curvatures are also quantitatively reproduced by the models. This is shown in the bottom right panel of Fig.\,\ref{fig:model_eps}, where we plot the radial modes of RGB (blue circles) and secondary clump (grey squares) for 2.5\,$M$\sun\ models with \dnu\ ranging from 8 to 9\mh\ (indicated by black symbols in the HR-diagram). For comparison we also plot the radial modes of 1.0\,$M$\sun\ RGB models. 
Even though these models do not represent our observed sample best (for that we would have to include 1.0\,$M$\sun\ RC models, which we cannot compute) the general trend is clear.
The mode curvature is largely defined by the model mass, although the evolutionary stage has some influence, shifting the lower order modes towards higher frequencies and the higher order modes towards lower frequencies when changing from the RGB to the secondary clump. The effect is small but results in a difference between central and average \dnu\ of about 0.5\,\% and is therefore accessible in the observations.

\begin{figure}[t]
	\begin{center}
	\includegraphics[width=0.5\textwidth]{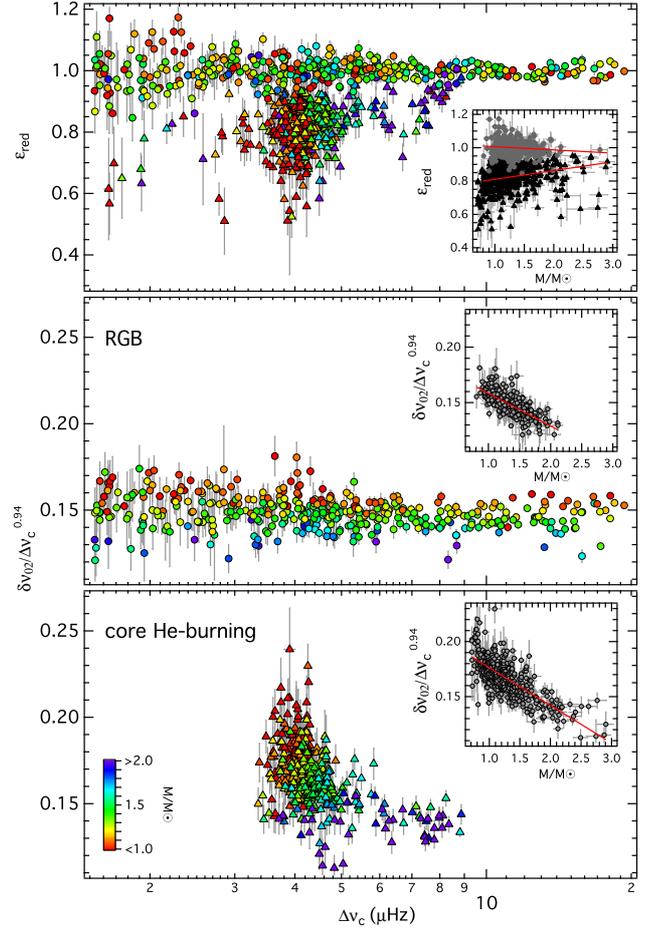}
	\caption{Reduced phase shift and central small frequency separation for RGB (circles) and core He-burning stars (triangles) with the symbol colours coded according to stellar mass. Inserts show the mass dependency of the various parameters along with a linear fit (red lines). In the insert of the top panel, RGB and RC stars are indicated by grey circles and black triangles, respectively. Note that stars with \dnuc\ below 1.5\mh\ are excluded here because their uncertainties are too large. Colours are given only in the online version.} 
	\label{fig:mass} 
	\end{center} 
\end{figure}

\section{The influence of the stellar mass}	

We have demonstrated that the various parameters we have chosen to describe the acoustic spectrum of red giants are affected by the evolution of the stellar core, but it can be expected that they also depend on other properties of the star. One of these is the stellar mass, which we can estimate via seismic measurements. Following the approach of Paper I, we compared the measured \num\ and \dnu\ to those of an extensive grid of stellar models, where the seismic parameters of the models were determined from scaling relations. A problem in the original approach was the ambiguous evolutionary stage of the stars, leading to a relatively uncertain position in the HR diagram in the region that is populated by both RGB and RC stars. Knowing the evolutionary stage of the individual stars, we can disentangle the contributions of RGB and RC models to the fit. This results in better constrained effective temperatures and luminosities but does not greatly affect the determination of the stellar mass (or radius). With the present observations, the mass is typically constrained to within about 5\,\% assuming that the seismic scaling relations are correct. This is not entirely true as can be seen from Fig.\,\ref{fig:model_dnu} \citep[or e.g.,][]{mig11} and an error of about 2\,\% in the \dnu\ scaling relation itself adds another about 4\,\% uncertainty, although a detailed error analysis is not of particular interest here and beyond the scope of this paper. 

In the top panel of Fig.\,\ref{fig:mass} we show the influence of the stellar mass on the phase shift. To correct for the relation between \epsc\ and \dnuc , we fitted a linear function to the RGB stars in the \epsc\ - \dnuc\ diagram and define $\epsilon_\mathrm{red} = \epsilon_c / (0.58+0.66 \log \Delta\nu_c)$. Our fit is in good agreement with that found by \cite{cor12} in their study of cluster red giants in the \textit{Kepler} field but slightly different from the relation adopted by \cite{mos11a} based on the shorter series from CoRoT. For RGB stars we find no significant correlation between $\epsilon_\mathrm{red}$ and $M$ (see the grey symbols in the inset), which is consistent with \cite{cor12} and to the study by \citet{whi11} in their extensive analysis of stellar model frequencies. The core He-burning stars show a slight correlation with $M$, which does not fully explain the dispersion in $\epsilon_\mathrm{red}$. The dispersion is significantly larger for RC stars (black symbols) than for RGB stars and can be as large as 30\,\% of \dnuc . This also explains why we do not find clear $l$ = 0 and 2 ridges in the right panel of Fig.\,\ref{fig:comp_echelle}. The explanation for this is unknown but chemical composition might be a good candidate. 
More interesting here is the gradient of increasing mass with increasing \dnuc , which results from the fact that all RC stars have a very similar radius (around 11\,R\sun , see Paper I).

The dependence of \dn2\ on the stellar mass is expected and has already been reported for red giants qualitatively from observations \citep{hub10} and from theoretical considerations \citep{mont10} but has not yet been quantified from actual measurements \citep[apart from the very recent study by][]{cor12}. From Fig.\,\ref{fig:eps} we see that \dn2\ does not scale linearly with \dnuc , which is presumably due the evolution of the burning H shell.  To account for this we fitted a power law to the RGB stars and found that a scaling with \dnuc$^{0.94}$ removes the evolutionary effect best. In the middle and bottom panels of Fig.\,\ref{fig:mass} we plot the normalised quantity $\delta\nu_{02}/$\dnuc$^{0.94}$ for RGB stars and core He-burning stars, respectively, with the symbol colour indicating the stellar mass. Both populations show a clear trend with $\delta\nu_{02}/$\dnuc$^{0.94}$ decreasing for increasing mass. The dependency is quantified by a linear fit $a + b \,M$ (see red lines in the insets), with the coefficients $a$ and $b$ being equal to 0.188 and $-$0.029 for the RGB stars and 0.210 and $-$0.034 for the RC stars, respectively, and $M$ in solar units. We note that the coefficients presented here differ from those determined by \cite{cor12}, which is because they assumed a linear relation between the large and small separation. However, if we compare our results with theirs we derive estimates for \dn2 that agree within a few per cent.

\section{Summary and conclusions}

When red giants evolve from stars burning H in a shell to stars that also burn He in their core, they largely reconfigure their interior structure but still populate the same region in the HR-diagram and are therefore difficult to distinguish. However, recent studies have shown that the mixed p/g dipole modes probe the stellar core and that their period spacings give a particularly clean separation of the two groups. In this paper we investigated the radial p-mode spectrum of red giants and how it is affected by the structural changes due to stellar evolution. Even though the radial modes do not probe the core itself, the p-mode cavity reaches deep into the star and at least its lower boundary should be sensitive to the evolution of the stellar core.

We have analysed high-precision photometric time series from the first 600 days of \textit{Kepler} observations for a large sample of red giants. We have measured the local large frequency separation $\pm$1 radial order around the centre of the power excess and determined the phase shift of the central radial mode from a local development of the asymptotic relation. Plotting this local pair of seismic observables (\dnuc , \epsc ) against each other reveals different populations of stars, with the RC and secondary clump stars having a systematically smaller \epsc\ than their counterparts on the RGB. There are also a few stars at low \dnuc\ which do not fit into the picture of RGB stars, which we identify as stars that have already left the RC to climb the AGB. A crosscheck with the mixed mode period spacings shows that our classification inferred only from the three central radial modes is as reliable as the mixed mode period spacing to distinguish between red giants in different evolutionary stages. 

We found a tight correlation between \epsc\ and \dnuc\ for RGB stars. It appears that the difference in \epsc\ between the RGB and clump stars becomes smaller and eventually indistinguishable if we use an average of several radial orders, instead of a local, i.e. only around the central radial modes, large separation to determine \epsc . With this phenomena apparently being caused by different large separations, the average \dnug\ turned out to be systematically smaller than the central \dnuc\ in RC stars whereas there is no significant difference for the RGB stars. Clearly, the information on the evolutionary stage is encoded locally, more precisely in the shape of the radial mode sequence. This turns out to be, on average, approximately symmetric around the central radial mode for RGB stars but asymmetric for RC stars. In the case of the symmetric sequence, one will always measure about the same \dnu\ as long as roughly the same number of modes are considered above and below the central radial mode. This is not so for the asymmetric sequence of RC stars, which results in different estimates of \dnu\ (that differ on average by about 1.5\,\%) depending on whether all modes or only the central modes are taken into account. It is also worth mentioning that neither an average \dnug\ nor the central \dnuc\ used in the present analysis is an ideal parameter since none of them is an exact representation of the asymptotic large spacing. However, according to stellar models, \dnuc\ is an at least as good and mostly better proxy for the asymptotic spacing than \dnug\ and should therefore be preferred in an asymptotic analysis of the observed frequencies. We computed radial mode frequencies for a sequence of red-giant models and found them to qualitatively confirm our findings for both the evolution of the phase shift of the central radial mode, as well as the with stellar evolution changing shape of the radial mode sequences. We also found a clear signature of the evolutionary stage in \dn2\ and quantified the mass dependency of this seismic parameter. 

Summarising, we found a reliable and relatively easy-to-access indicator for the evolutionary stage of red giants based only on the three central radial modes. This is of particular interest for red giants with very weak dipole modes \citep[e.g.,][]{mos11c} or bright red giants, which are not accessible to \textit{Kepler} (or CoRoT) and for which we cannot resolve the mixed mode spectrum in the near future and therefore measure their period spacing. One of them is $\epsilon$ Oph, which is a bright and well studied red giant \citep[e.g.,][]{kal08a}. After the initial MOST \citep{walk03} observations in 2005, $\epsilon$ Oph has been re-observed twice \citep{kal11} so that the data are now sufficient to accurately measure the phase shift of the central radial mode and to identify the star as a rather massive secondary clump star (Kallinger et al. in preparation), which is an important input for subsequent modelling.

\begin{acknowledgements}
The authors gratefully acknowledge the \textit{Kepler Science Team} and all those who have contributed to making the \textit{Kepler} mission possible. Funding for the \textit{Kepler Discovery mission} is provided by NASA's Science Mission Directorate. TK and JDR are supported by the FWO-Flanders under project O6260 - G.0728.11. SH acknowledges financial support from the Netherlands Organisation for Scientific Research (NWO)

\end{acknowledgements}

\bibliographystyle{aa}
\bibliography{epsilon}

\end{document}